# MEASUREMENTS OF SCRF CAVITY DYNAMIC HEAT LOAD IN HORIZONTAL TEST SYSYTEM

B. D. DeGraff, R. J. Bossert, L. Pei, W. M. Soyars

Fermi National Accelerator Laboratory
Batavia, IL, 60510, USA

## ABSTRACT

The Horizontal Test System (HTS) at Fermilab is currently testing fully assembled, dressed superconducting radio frequency (SCRF) cavities. These cavities are cooled in a bath of superfluid helium at 1.8K. Dissipated RF power from the cavities is a dynamic heat load on the cryogenic system. The magnitude of heat flux from these cavities into the helium is also an important variable for understanding cavity performance. Methods and hardware used to measure this dynamic heat load are presented. Results are presented from several cavity tests and testing accuracy is discussed.

## INTRODUCTION

The Horizontal Test System (HTS) at Fermilab is part of the continuing research and development program to produce superconducting radio frequency (SCRF) cavities at various frequencies and for various projects [1]. The purpose of the HTS is to test fully dressed cavities in the horizontal orientation after they have been certified by vertical testing at other facilities at Fermilab. The HTS test cavities are inside the Horizontal Test Cryostat (HTC) [2], which is connect to the main Meson Detector Building (MDB) cryogenic supply system.

Operation of the HTS occurs at a range of saturated liquid temperature between 1.7 K and 2.0 K depending on the physics needs. Liquid helium is delivered to the HTS at 4.7 K and 2.48 x $10^5$ Pa (21 psig). Operations at the desired temperatures require active warm pumping on the liquid helium bath. This is accomplished by a Kinney roots blower (Model KMBD 10,000) and a Kinney liquid ring pump (Model KLRC 2100).

Several measurements are made of cavities that are tested in the HTS. One of these very important measurements is the dimensionless $Q_o$ quantity. The $Q_o$ of the cavity is measured very precisely when the bare cavity is tested in the vertical orientation. Due to the beam requirements for any future accelerator that will use either $1.3 \times 10^9$ Hz or $3.9 \times 10^9$ Hz cavities, the dressed cavities being tested in the horizontal orientation must be over-coupled, making any direct measurement of the cavity's $Q_o$ impossible. The alternative to a direct measurement of the $Q_o$ is to measure the dynamic heat load of the cavity and from there calculate the $Q_o$. The relationship between the $Q_o$ of the cavity and the dynamic heat load transferred to the superfluid helium bath is governed by equation (1).

$$Q_{DHL} = \frac{DF \cdot E^2_{ACC} \cdot L^2_{CAV}}{\left(\frac{R}{Q}\right) \cdot Q_0} \tag{1}$$

Where $Q_{DHL}$ is the cavity dynamic heat load; DF is the duty factor; $E_{ACC}$ is the cavity's accelerating gradient; $L_{CAV}$ is the cavity's length; and the ratio (R/Q) is a measure of cavity current.

## DYNAMIC HEAT LOAD TESTING

### Fixed Mass Flow Rate Testing

When the original plan was proposed for a cavity test facility in MDB, the need for precise dynamic heat load measurements was not addressed. The warm vacuum pump skid had a flowmeter installed on the discharge header, but the flowmeter was sized based on pump capacity and not based on cavity testing needs.

Several attempts were made both at the CC2 installation and at the HTS installation to measure the dynamic heat load of the cavity using a fixed mass flow rate method. This was accomplished by fixing the position of the Joule-Thompson (JT) valve that normally was used to regulate the level of the bath surrounding the cavity. The valve was fixed at a position slightly higher than normal operating range, and then the cartridge heater located next to the cavity was turned on and adjusted until a steady liquid level was observed. When power was introduced to the cavity, the power supplied by the heater was decreased in an attempt to maintain a stable liquid level.

This method was successful in providing a high limit for the actual dynamic heat load of the cavity. The problem with this approach is that the quality of helium flow rate through the JT valve is strongly influenced by slight variations in inlet temperature. Variations of +/- 0.05 K on the inlet temperature to the JT valve contribute to a total uncertainty in cavity dynamic heat load with this method of +/- 0.5 W.

### Precision Testing Equipment Used

In an attempt to provide a more refined measurement of the cavity dynamic heat load, a low range flowmeter was installed on the discharge header of the vacuum pump. This new flow meter is a FCI (Model GF90) thermal mass flow meter with a range of 0 to 0.0012 kg/s. The accuracy of the FCI flowmeter is described by equation (2).

$$Error = 0.01 \cdot Reading + 0.005 \cdot FullRange \tag{2}$$

While this error is not insignificant when attempting to resolve small changes, it was a substantial improvement over the old flow meter.

**TABLE 1.** Heater Power Measurement Accuracy.

| Applied Heater Power (W) | Accuracy (W) | |
|---|---|---|
| | HTC | CC2 |
| 1 | +/- 0.001 | +/- 0.021 |
| 3 | +/- 0.004 | +/- 0.032 |
| 5 | +/- 0.014 | +/- 0.044 |

The accurate measurement of cartridge heater power applied to the cavity helium bath is also very important for verifying the calculation of cavity dynamic heat load. These heater measurements done for both the CC2 and the HTC heaters were accomplished by measuring the voltage with a Fluke (Model 179) True RMS multimeter. The current was measured using a MetraHit (Model 29S) multimeter. Both had current calibration seals intact. The Metrahit meter was used for the current reading since it is regarded as one of the most accurate hand held meters available. TABLE 1 shows the accuracy for different heater power calculations at CC2 and HTS. The difference in accuracy is due to the cartridge heaters having different resistance values.

## Verification of Methodology in the Capture Cavity II Cryostat

Testing first occurred in the CC2 cryostat since the HTS was unavailable. A control volume analysis was applied around the cavity, taking temperature and pressure measurements at both the inlet and exhaust of the cavity piping. The flow rate used for these calculation was recorded by the FCI flow meter described above. All measurements were averaged over a period of no less than 0.5 hours in order to ensure that normal system oscillations had a minimal effect on the data.

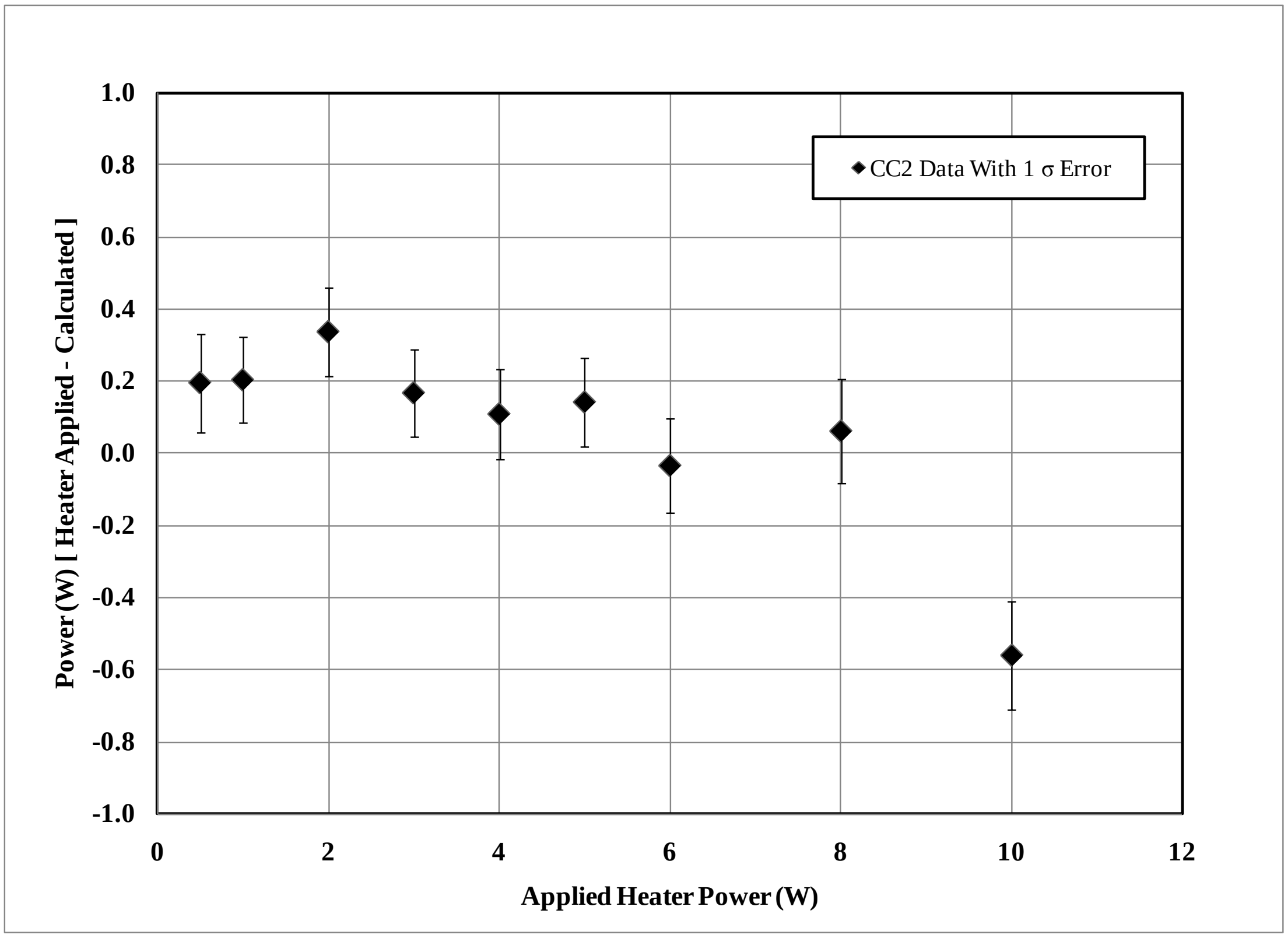

**FIGURE 1.** Results of CC2 Heat Load Testing for Various Heater Settings.

Cavity dynamic heat load for these tests were simulated by using the cartridge heater with the voltage and current measurements described above. A baseline measurement was taken with no heater power. Power was then applied to the heater. The difference in the two measurements is the calculated heat added to the liquid helium bath. Testing at CC2 occurred at bath pressure of 1600 Pa (12 Torr). FIGURE 1 show the results of CC2 testing for various applied heater settings. There is good agreement between applied and calculated power. The anomaly that occurred at 10 W is most likely due to the lack of loop tuning.

The error analysis included in FIGURE 1 is the result of collecting the raw data standard deviations. The error was then propagated through the calculations using a root mean square technique [3]. The y-axis error bars are 1 sigma error. The x-axis error bars are not displayed as they are too small to be of significance as described above in TABLE 1.

## Verification of Methodology in the Horizontal Test System Cryostat

The testing in the HTS follow the same methodology as in CC2. The same equipment was used to measure temperature, pressure, voltage, current, and flow rate. The results of testing in HTS using the cartridge heater are displayed in FIGURE 2. Testing was repeated multiple times, with approximately the same results. The calculated heater power is consistently higher than the applied heater power. A polynomial regression was fit to the data and shown as Equation (3).

$$y = -0.025 \cdot x^2 - 0.294 \cdot x - 0.053 \tag{3}$$

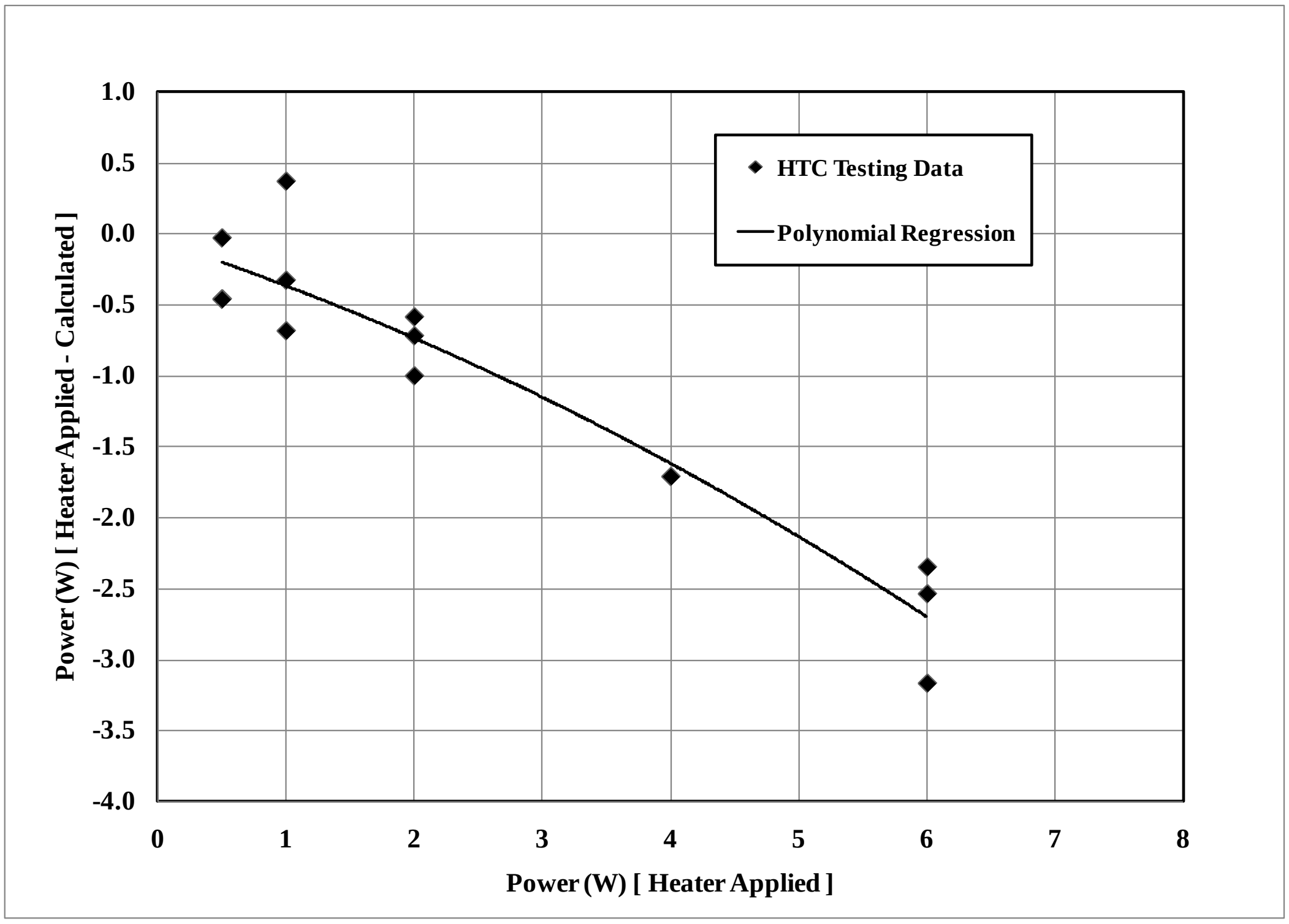

**FIGURE 2.** HTC Heater Testing Data with Trend Curve.

The polynomial curve fit does a good job representing the data with a $R^2$ value of 0.906. Several theories have been postulated to explain these results, but none have shown to explain this observation. Error calculations were also done for the data shown in FIGURE 2 and they are of the same magnitude as those calculated during the CC2 tests. Data collected at 6 W fall outside of three sigma of statistical error.

### 3.9 GHz Cavity Data in the HTS

Data was collected from the 3.9 x $10^9$ Hz cavity inside the HTS cryostat. No cartridge heater power was applied during these tests. The cavity was run at different gradients, and the dynamic heat loads were calculated using the same methodology and hardware as the CC2 and HTS verification tests. The data shown in FIGURE 3 is not corrected by the regression equation derived from the HTS verification testing.

The cavity testing seems to be in good agreement with the predicated heat loads based on an assumed $Q_o$. Data for values of $Q_o$ equal to both 1 x $10^{10}$ and 2 x $10^9$ are also shown on FIGURE 3. The data indicated that the cavity has a $Q_o$ value of 2 x $10^9$.

## CONCLUSIONS

Accurate measurement of the dynamic heat load at both the CC2 and HTS test caves is possible. Data from these tests can be used to accurately map the $Q_o$ value of the cavity being tested. Future tests will be needed to study the strange discrepancy between the calculated and applied heater power observed in the HTS cryostat. Possible upgrades to the cartridge heater and thermal insulation may shed some light on the problem, but currently no schedule exists for these upgrades.

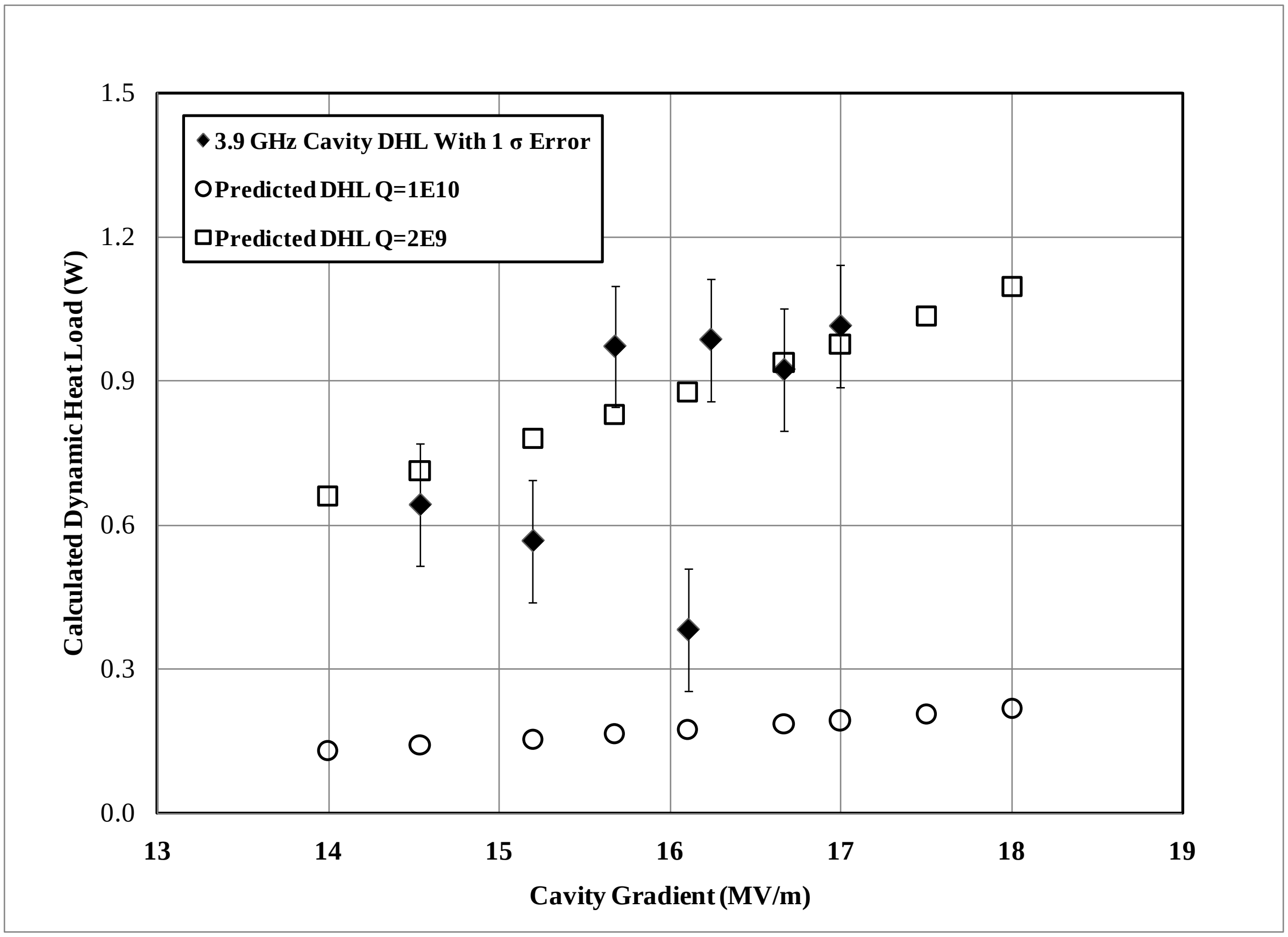

**FIGURE 3.** HTC Cavity Testing Data with Qo Curves.

Fermilab is preparing to go into production of 1.3 x $10^9$ Hz cavities for future ILC cryomodules and the HTS will be able to provide corrected dynamic heat load measurements to within +/- 0.2 W.

## ACKNOWLEDGEMENTS

Fermilab is operated by Universities Research Association Inc. under Contract No. DE-AC02-76CH03000 with the United States Department of Energy. The authors are grateful to all those in the Fermilab Cryogenic Department who have contributed to the testing and analysis of the MDB flowmeter. Special thanks to Andy Hocker and Elvin Harms for their efforts in making dedicated flowmeter testing time available.